\documentclass[twocolumn,showpacs,preprintnumbers,amsmath,amssymb]{revtex4}
\usepackage{graphicx}
\begin{document}
\title{In-medium $T$ matrix for neutron matter}
\baselineskip=1. \baselineskip

\author{P. Bo\.{z}ek}
\email{piotr.bozek@ifj.edu.pl}
\author{P. Czerski}
\email{czerski@horus.ifj.edu.pl}
\affiliation{Institute of Nuclear Physics, PL-31-342 Cracow, Poland}

\date{\today}





\begin{abstract}
We  calculate  the equation of state of pure neutron matter, comparing 
the $G$-matrix calculation with the in-medium $T$-matrix result.
At low densities, we obtain  similar  energies per nucleon, however some 
differences appear at higher densities.
We use the self-consistent spectral functions from the $T$-matrix approach
 to calculate the 
$^1S_0$ superfluid gap including self-energy effects.
We find a reduction of the superfluid gap by  $30\%$.

\end{abstract}





\maketitle

In the interior of neutron stars, neutron rich nuclear matter is present in
a range of densities up to several times the normal nuclear density.
Two properties of the neutron matter are important for the modeling of
neutron stars.
The equation of state, which serves as an input for the calculation of the
mass of the star, and the superfluid parameters of the dense  matter, 
relevant for the cooling and the rotation of stars.
Dense neutron matter cannot be studied directly in laboratory; therefore,
it is important to have a reliable description of this many-body state from
theoretical calculations.
Very advanced methods based on the Brueckner-Hartree-Fock (BHF) 
approximation were
applied for the calculation of neutron matter properties
\cite{baldonstar}. Independently, 
approaches using  the variational chain summation
to tackle  the
nuclear many-body problem
\cite{vcs1,vcs2} have been developed to a high level of accuracy. 
For symmetric nuclear matter, these calculations  have shown the need to
introduce a three-body force to reproduce the phenomenological saturation
point.  Relativistic corrections have also been estimated \cite{vcs2}.
Variational and BHF methods allow to study the equation of state of
neutron matter up to densities several times the normal nuclear matter
density.
Microscopic calculations of different pairing gaps in
neutron matter are also available 
\cite{lombardonstar,tatsuga,kkc}.

An alternative many-body scheme for the calculation of the properties of
nuclear matter is the self-consistent in-medium $T$ matrix
\cite{di1,Bozek:1998su,Dickhoff:1999yi,Dewulf:2000jg,Bozek:2001tz}
\begin{equation}
T=V+ VGGT 
\end{equation}
where the Green's functions $G^{-1}= {\omega-p^2/2m -\Sigma}$ 
in the ladder propagator
are dressed
by the self-consistent self-energy
\begin{equation}
i\Sigma= Tr[T_AG]  \ .
\end{equation}
The $T$-matrix approach is a conserving approximation 
\cite{Baym} and gives consistent
single-particle properties.
The details of the equations and of the solution in the real time formalism
can be found in \cite{Bozek:2002em,Bozek:1998su}.
  Since the self-energy has a non-zero
imaginary part, we obtain in the $T$-matrix approximation a nontrivial
single-particle spectral function.

At low densities, the $T$-matrix ladder diagrams are dominated by the
particle-particle propagators and therefore should reduce to the $G$-matrix 
result (if the influence of the nucleon dressing is small at low densities).
However, at higher densities the $\Phi$-derivable $T$-matrix approximation
takes into account some diagrams of higher order in the hole-line expansion.
It is then  instructive to compare the results of the $T$-matrix calculation
with the corresponding BHF calculation for a realistic 
interaction.

\begin{figure}
\includegraphics[width=0.48\textwidth]{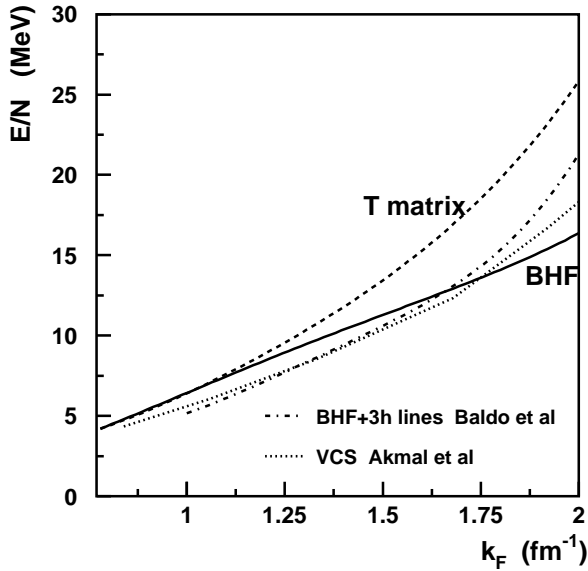}
\caption{\label{befig}  The energy per nucleon in pure
 neutron matter as a function of
  the Fermi momentum for the $T$-matrix (dotted line) and the BHF
(solid line) calculations using the Paris potential. The dotted line
represents the results of a variational chain summation \cite{vcs2}
and the dashed-dotted line the results of a BHF calculation with 3 
hole-lines contribution \cite{3holesnm}, 
both for a two-body Argonne V18 interaction.}
\end{figure}

We  perform self-consistent iterations of the $T$-matrix equations in
pure neutron matter, using the separable Paris interaction 
\cite{parisseparable,parisseparable2} and the  same methods as in Ref. 
\cite{Bozek:2002em}. In Fig. \ref{befig} the resulting 
energy per nucleon is plotted 
as a  function of the Fermi momentum. It is compared to the
BHF calculation with the continuous choice of the single particle energy,
 using the same interaction and numerical grids.
An excellent consistency of the two methods is found at low densities,
as expected. This provides a powerful check of the methods used for the
solution of integral equations, which for the $T$-matrix case involve 
refined numerical algorithms. At higher densities the neutron matter in
the $T$-matrix calculation is less bound, leading to a considerably harder
equation of state. It is known that the BHF results with the continuous choice
are not modified considerably by  $3$ hole-lines corrections
\cite{3holes,3holesnm}. The $T$-matrix approximation is a different summation
of some of the $3$ and more hole-lines diagrams. We obtain the first 
quantitative estimate of the difference between the self-consistent $T$-matrix
approximation and the BHF approach. Since the latter one is believed to be 
close to the converged result, we find an estimate of the accuracy of the
$T$-matrix calculation. Below $k_F=1.3 \textrm{fm}^{-1}$ 
the difference in the energy per particle from  the
two calculations is less than $2\textrm{MeV}$. The discrepancy at higher
densities could be attributed to the ring diagrams contribution.  
Also at higher densities meaningful
 results can be obtained only after inclusion
of higher partial waves and  three-body forces effects.
In Fig. \ref{befig}, results of modern variational
calculations \cite{vcs2} and of a BHF calculation including
contributions from 3 hole-lines diagrams \cite{3holesnm} are also shown.
Both methods are using  a two-body 
Argonne V18 potential and give similar results. As
shown in Ref. \cite{3holesnm} the continuous BHF results are almost
indistinguishable from the full BHF calculation with 3 hole-lines
contributions included.
Therefore, the difference between  our BHF result (solid line in
Fig. \ref{befig})
and the variational or next order BHF
 results must be attributed to a different interaction used,
and to the limited number of partial waves taken.
The $T$-matrix energy per particle is larger than obtained
in other approaches. Most importantly it is significantly larger
than the continuous BHF calculation with the same interaction. 
It indicates that 
the $T$-matrix calculation does not give the correct  energy per particle at
higher densities.

For the analysis of  superfluidity, 
we use the spectral function obtained in the $T$-matrix
calculation. As noted above, the $T$-matrix approach gives reliable results
for the single-particle properties in the  medium. 
The effective mass in our calculation decreases from $1.07$ 
to $1.02$ nucleon masses with increasing density, which is 
only about $0.05$ more than
the corresponding BHF effective mass. 

The superfluid gap equation with full spectral functions is
\begin{eqnarray}
\label{gapfull}
\Delta(p)=\int\frac{d \omega d\omega^{'} d^3k}{(2\pi)^5} A(k,\omega) 
A_s(k,\omega^{'}) \nonumber \\
\frac{\left[1-f(\omega)-f(\omega^{'})\right]}{-\omega-\omega^{'}}
V(p,k)\Delta(k)
\end{eqnarray}
where $A_s(p,\omega)$ denotes the spectral function of the
nucleon, including the diagonal self-energy $\Sigma(p,\omega)$ 
(obtained in the $T$-matrix
 approximation) and the off-diagonal self-energy $\Delta(p)$ (obtained from
Eq. (\ref{gapfull}) itself); 
$A(p,\omega)$ is the spectral function of the nucleon dressed
with the  diagonal self-energy only. 
Self-energy effects in the gap equation were studied in a number of 
papers \cite{Baldo:2000zf,Bozek:2000fn,Lombardo:2001vp,Bozek:2002jw}.
It was found that generally the superfluid energy gap is reduced by the 
in-medium dressing of nucleons.
The mechanism of this reduction can be understood easily from the effective
quasiparticle gap equation
 \cite{Bozek:2000fn,Lombardo:2001vp}
\begin{eqnarray}
\label{gapz} & &
\hat{\Delta}(p)= \nonumber \\ 
& & -\int\frac{d^3k}{(2\pi)^3}Z_kZ_p V(p,k)
\frac{\left[1-2f(E_k)\right]}
{2E_k}\hat{\Delta}(k) \ 
\end{eqnarray}
which is obtained as the quasiparticle limit of the gap equation with dressed
propagators (\ref{gapfull}).
The superfluid energy gap 
is 
\begin{equation} 
\label{gape}
\hat{\Delta}(p)=Z_p\Delta(p)
\end{equation}
 where 
\begin{equation}
Z_p=\left(1-\frac{\partial \Sigma(p,\omega_p)}{\partial
    \omega}|_{\omega=\omega_p}\right)^{-1} \ 
\end{equation}
and
\begin{equation}
 E_p=\sqrt{(\omega_p-\mu)^2+\hat{\Delta}(p)^2} \ .
\end{equation}
The usual BCS equation is obtained by putting $Z_p=1$ in the above
equations. In that case, the only influence of the medium comes through the
modification of the dispersion relation $\omega_p$.

Dispersive self-energy corrections to the gap equation 
are twofold. First, the 
superfluid energy gap is multiplied by the quasiparticle strength $Z_{p_F}<1$
(\ref{gape}),
and second, the interaction strength in the gap equation
is reduced by a factor $Z_pZ_k$ (\ref{gapz}).
In the case of symmetric nuclear matter at normal density 
the reduction of the superfluid gap
due to self-energy effects is of about one order of magnitude 
\cite{Bozek:2002jw}. 

\begin{figure}
\includegraphics[width=0.48\textwidth]{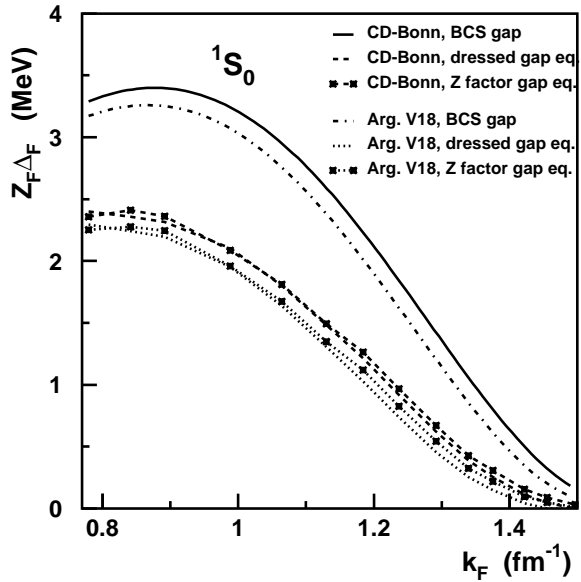}
\caption{\label{gapfig}  The $^1S_0$ superfluid energy gap in neutron matter,
  for the CD-Bonn interaction in the BCS approximation (solid line), from the
  gap equation with dressed propagators (\ref{gapfull}) (dashed line),
 and from the effective   gap equation (\ref{gapz}) (dashed line with stars);
and for the Argonne V18 interaction 
in the BCS approximation (dashed-dotted line), from the
  gap equation with dressed propagators (\ref{gapfull}) (dotted line),
 and from the effective   gap equation (\ref{gapz}) (dotted line with stars).}
\end{figure}

In Fig. \ref{gapfig} we present the results of an analogous calculation for
the $^1S_0$ pairing in neutron matter for two interaction potentials
\cite{cdbonn,v18}. 
It is known that for the $^1S_0$
gap different interaction potentials lead to similar superfluid gaps
\cite{lombardonstar}.
We recall  that the spectral functions and the single-particle energies 
used in the gap equation are obtained from a separate  
self-consistent calculation in the normal phase.
We find the $Z_p$ factor at the Fermi
momentum $\simeq 0.9$ 
for the range of densities studied. It is close to the one
obtained from the BHF calculation without rearrangement terms 
\cite{Lombardo:2001vp}. The rearrangement terms in the BHF approach 
reduce this value further \cite{Lombardo:2001vp}
(the $T$-matrix result includes the rearrangement
contribution to the self-energy).
We find a smaller $Z$ factor than
the authors of reference \cite{Baldo:2000zf}. 
The obtained values of $Z_p$ translate directly into the resulting superfluid
gap, which we find in between the results of
Refs. \cite{Baldo:2000zf} and \cite{Lombardo:2001vp}.
The average reduction factor is  $0.7$ in the region of the maximal gap.
 The effective gap equation 
(\ref{gapz}) is an excellent approximation to the solution of the
gap equation with full spectral functions (\ref{gapfull}).
It is due to the fact that in this region of densities, the gap equation 
kernel is
dominated 
by contributions close to the Fermi momentum, where the quasiparticle 
limit applies.
We could not perform the self-consistent $T$-matrix calculation at
densities below $0.016\textrm{fm}^{-3}$; 
the expected disappearance of  self-energy effects in  the gap equation at low
densities could not be explicitly demonstrated. This is in contrast
to the results for the energy per particle where the BHF approach
and the self-consistent $T$-matrix calculation merge at low
density \cite{fw} (Fig \ref{befig}).

In the gap closure region the ratio of the BCS gap to
the gap obtained with dressed propagators is large. 
Such a strong reduction of the gap was found in symmetric nuclear matter at
normal density \cite{Bozek:2002jw}, 
which is  close to gap closure for the $^3S_1-^3D_1$ gap.
 Also the effective gap equation 
(\ref{gapz}) and the gap equation with full spectral functions (\ref{gapfull})
give markedly different results.

In summary, we study pure neutron matter using  the in-medium $T$-matrix
approximation.
 We  present a comparison of the energy per nucleon for neutron matter
in the BHF 
 and $T$-matrix approaches. We find  similar binding
 energies at low densities, but a stiffer equation of state for the $T$-matrix
 calculation at higher densities. This is an explicit confirmation of the
 expected similarities and discrepancies between the two approaches.

We  use  self-consistent spectral functions in neutron matter to
calculate the superfluid gap from the gap equation with dressed propagators.
This allows us to test the accuracy of the effective gap equation
(\ref{gapz}), which incorporates to some extend  self-energy effects.
 Using full spectral
functions, we find, in the maximal gap region, a reduction of the
gap of about $30\%$ and a very good agreement with the effective gap equation.
  In the gap closure region, we find a stronger
 reduction. In this region the effective gap equation (\ref{gapz})
is a less reliable 
estimate of  self-energy effects. It should be remembered that also other 
in medium effects modify the  superfluid gap
in neutron matter \cite{clarkpol,Wambach:1993ik,schulzepol,pol2}.

\acknowledgments
This work was partly supported by the KBN
under Grant No. 2P03B02019.

\bibliography{../mojbib}

\end{document}